# Growth-Induced Strain in Chemical Vapor Deposited Monolayer MoS$_2$: Experimental and Theoretical Investigation

*Satender Kataria,[a,*] Stefan Wagner,[a] Teresa Cusati,[b] Alessandro Fortunelli,[c] Giuseppe Iannaccone,[b] Himadri Pandey,[a] Gianluca Fiori[b] and Max C. Lemme [a*]*

[a]Graphene based Nanotechnology, School of Science and Technology, University of Siegen, Hölderlinstr. 3, Germany

[b]Dipartimento di Ingegneria dell'Informazione, Università di Pisa, Via G. Caruso, 16, Pisa 56122, Italy

[c]CNR-ICCOM, Consiglio Nazionale delle Ricerche, via Giuseppe Moruzzi 1, 56124, Pisa, Italy

**Abstract**

Monolayer molybdenum disulphide ($MoS_2$) is a promising two-dimensional (2D) material for nanoelectronic and optoelectronic applications. The large-area growth of $MoS_2$ has been demonstrated using chemical vapor deposition (CVD) in a wide range of deposition temperatures from 600 °C to 1000 °C. However, a direct comparison of growth parameters and resulting material properties has not been made so far. Here, we present a systematic experimental and theoretical investigation of optical properties of monolayer $MoS_2$ grown at different temperatures. Micro-Raman and photoluminescence (PL) studies reveal observable inhomogeneities in optical properties of the as-grown single crystalline grains of $MoS_2$. Close examination of the Raman and PL features clearly indicate that growth-induced strain is the main source of distinct optical properties. We carry out density functional theory calculations to describe the interaction of growing $MoS_2$ layers with the growth substrate as the origin of strain. Our work explains the variation of band gap energies of CVD-grown monolayer $MoS_2$, extracted using PL spectroscopy, as a function of deposition temperature. The methodology has general applicability to model and predict the influence of growth conditions on strain in 2D materials.

Since the discovery of graphene, two-dimensional (2D) materials have become the most intensively studied class of materials in a short span of around 12 years.[1,2] The enormous research interest in these materials is due to their distinct physical and chemical properties in the atomically thin regime. Among 2D materials, transition-metal dichalcogenides (TMDs) are the focus of current research on layered materials as they cover a whole range of interesting properties including metallic, semiconducting and superconducting.[3] Most notably, TMDs based on molybdenum (Mo) and tungsten (W) have attracted the research community's attention, due to their semiconducting nature with a wide range of band-gaps (in contrast to gapless graphene [3,4]), that makes them suitable for logic applications in electronics and optoelectronics.[5–11] In particular, the bandgap of molybdenum disulphide ($MoS_2$) exhibits a variation with the number of layers, i.e., it is indirect (~1.2 eV) in bulk form and becomes direct (~1.9 eV) in the monolayer limit.[12] The presence of a direct band gap and of a unique 2D structure of monolayer $MoS_2$ makes it an attractive material for novel applications. Monolayer $MoS_2$-based field effect transistors exhibit high on/off ratios of around $10^4 - 10^8$ with mobility on the order of $10^{-1}$ to $10^2$ $cm^2V^{-1}s^{-1}$.[13–17] Logic applications such as integrated circuits,[18] small-signal amplifiers with a room temperature gain of 4,[19] inverters and NAND gates[20] based on monolayer $MoS_2$ have also been demonstrated, paving the way for low-power nanoelectronics. Numerous other applications include chemical and biosensing,[21,22] photodetection,[23],[24] and flexible electronics.[25,26]

High quality and crystalline monolayer $MoS_2$ films can now be synthesized on a large-scale on a range of technologically relevant substrates using chemical vapor deposition (CVD).[27,28] In general, molybdenum trioxide ($MoO_3$) and sulphur (S) powders, are the most widely used reactants for CVD growth of $MoS_2$ layers. The most important aspect of CVD $MoS_2$ is that it can be synthesized directly on insulating substrates like silicon dioxide (SiO$_2$),[14,16,29,30] sapphire,[31] quartz[32] and hexagonal boron nitride.[33] Recently, the growth of polycrystalline monolayer MoS$_2$ on 4 inch wafers has been demonstrated using CVD based on metal organic precursors, opening the possibility to fabricate electronic devices based on MoS$_2$ on a large scale.[29] Generally, the CVD growth of MoS$_2$ proceeds via formation of isolated triangular grains, which then merge together to form continuous films of MoS$_2$. However, it is observed that different shapes and sizes of MoS$_2$ crystals can be formed on a substrate which points towards different stoichiometries of the crystals due to the variations in the precursor concentration.[34] The deposition pressure and amount of S precursor is also found to affect the morphology of monolayer MoS$_2$ grains which varies from being triangular, star-shaped, and truncated triangles.[16] To date, monolayer MoS$_2$ has been synthesized in a wide range of temperatures from around 600 °C to 1000 °C using the CVD technique,[15,16,27,34–37] but it has been found that CVD grown MoS$_2$ is prone to have intrinsic non-uniform tensile strain due to substrate interactions.[38] This suggests that MoS$_2$ grown at different temperatures may have different levels of strain, which, in turn, can affect the electronic and optical properties of the as-grown material. In addition, how the substrate affects the strain levels in MoS$_2$ is not yet clear. Therefore, it is worth to investigate and systematically compare the properties of monolayer MoS$_2$ synthesized at different temperatures, thus providing important information and guidelines for targeted material growth otherwise scattered and scarce.

In this work, we investigate the effect of deposition temperature on the growth of MoS$_2$ using CVD technique and present a comparison of their morphologies and optical properties. Detailed optical and SEM studies reveal the formation of different structures on the substrates and the same is explained on the basis of MoO$_{3-x}$ vapor flux arriving at a specific location on the substrate. Micro-Raman and photoluminescence (PL) studies indicate that monolayer MoS$_2$ grown at 650 °C has different optical quality due to different strain levels than the films grown at 700 °C, irrespective of their similar triangular morphology. We use density functional theory to examine the effect of homogeneous strain on the band gap energy of monolayer MoS$_2$ and investigate the role of the substrate during MoS$_2$ growth. Our work shows that growth temperature has a significant effect on the quality of CVD grown 2D TMDs, which eventually influences the performance of the devices based on these materials.

The MoS$_2$ monolayers examined in the present study were synthesized using solid MoO$_3$ and S precursors at atmospheric pressure. The synthesis conditions are detailed in the Experimental section. The deposition was carried out at three different temperatures, namely 600 °C (T1), 650 °C (T2) and 700 °C (T3). Figure 1 shows the surface morphology at the substrate location, which was directly above the precursor during the depositions. For sample T1, no significant deposits were observed, as seen in Figure 1a, except the formation of vertical nanostructures (Figure 1b). For samples T2 (650 °C) and T3 (700 °C), the size and density of these nanostructures were found to increase, indicating the formation of thicker films (Figures 1c-1f). This can be explained on the basis of higher vapor flux of evaporated material with increase in the deposition temperature. The formation of MoS$_2$ using solid MoO$_3$ and S precursors proceeds with the reduction of MoO$_3$ in a volatile species MoO$_{3-x}$, which then deposits on the substrate, and acts as a nucleation site for MoS$_2$ growth. The possible reaction steps are the following[39]:

$$MoO_3 + {}^{x}/_{2}S \rightarrow MoO_{3-x} + {}^{x}/_{2}SO_2, \quad (1)$$

$$MoO_{3-x} + {}^{(7-x)}/_{2}S \rightarrow MoS_2 + {}^{(3-x)}/_{2}SO_2. \quad (2)$$

The vapor flux of evaporated material, in this case $MoO_{3-x}$, is highest at substrate locations directly above the precursor. Along the length of substrate, there is a varying concentration of $MoO_{3-x}$ which decreases in the downstream direction, as discussed later. We observed that vertical nanostructures were diminishing as well as growth of lateral structures with different densities dominated towards the downstream end of the substrate (Figure S1). Similar observations have been made by Vila *et al*. where the growth of vertical $MoS_2$ nanosheets along with the lateral growth on different substrates is reported and the same has been explained on the basis of $MoO_x:S_2$ partial pressure and its gradient across the substrate length.[40]

Figure 2a and 2b show the photographs of sample T2 and T3, respectively. It is evident that more deposition has occurred on sample T3 as compared to T2. Figure 2 shows the surface morphology of $MoS_2$ films or grains towards the substrate edges (inside the area depicted by red ovals in Fig. 2a and 2b) for samples T2 and T3. We did not observe such grain formation on sample T1 under the present deposition conditions. These grains, as confirmed by Raman and PL measurements and discussed later, are mostly monolayers. Electrical measurements conducted on continuous film regions revealed the typical n-type nature of the as-deposited monolayer $MoS_2$ films (Figure S2). Figures 2a and 2c show the optical and SEM micrographs of the grains on sample T2. Along with the $MoS_2$ layers, small nanoparticles of less than 100 nm size were also deposited. Sample T3 showed the same grain morphology towards substrate edges, as shown in Figure 2b and 2d. However, nanoparticles of different shapes, with dimensions above 200 nm, were also deposited. These nanoparticles seemed to act as nucleating sites for adlayers (on top of the grains) and for new layers on the substrates, seen as dark contrast under the nanoparticles in Figures 2c and 2d. These changes in morphologies or structures can be explained with an increased flux of $MoO_{3-x}$ arriving at the substrate at higher deposition temperatures. Therefore, different strategies like MoO$_3$ nanoribbons,[16] proximity effects,[41] and low flow rates of metal organic precursors are adopted to obtain uniform and continuous deposition of monolayer MoS$_2$ films.[29] In the present methodology, which is the most conventional one, substrates are kept directly above the precursor with face down position. Therefore, the substrate position just above the precursor is directly exposed to the evaporant flux, resulting in the formation of thicker films and different nanostructures i.e. vertical ones. As one move towards the substrate edges, the evaporant flux decreases, resulting in an unintentional reduced supply of MoO$_{3-x}$ species at these places. This reduced evaporant flux or concentration can be described with a cosine dependence[42]:

$$A \propto cos(\theta)cos(\beta)/d^2, \qquad (3)$$

where, $A$ is the rate of vapor flux arriving on the substrate, $\theta$ is the angle between the normal to the substrate and the direction of vapor flux, $\beta$ is the incidence angle of vapor flux on the substrate, and $d$ is the distance between the center of the substrate and the vapor source. In our case, the distance between vapor source, i.e. MoO$_3$ powder, and the substrate is constant. However, vapor flux arriving at a substrate may be different at different locations on the substrate as per Eq. 3, with maximum flux arriving directly above the precursor ($\theta = \beta = 0°$). The locations near substrate edges have higher $\theta$ and $\beta$, which lead to lower concentrations of MoO$_{3-x}$ species. This creates suitable growth conditions for the formation of monolayer MoS$_2$ films or isolated grains, as observed in the present study.

We studied the optical properties of as-grown monolayer MoS$_2$ grains using micro-Raman and PL spectroscopy. MoS$_2$ samples deposited at different temperature exhibited quite distinct optical properties even though they had similar morphology. Since, we did not observe triangular grains on Sample T1, we compare the optical properties of samples T2 and T3, where isolated triangular grains were mainly observed. For this, we chose grains without adlayers, as optical properties of the 2D materials exhibit observable changes with the number of layers. Figure 3 shows the Raman spectra of isolated $MoS_2$ grains on samples T2 and T3. In general, Raman spectra of $MoS_2$ show two characteristic modes, the $E^1_{2g}$ mode, corresponding to in-plane vibration of Mo and S atoms, and the $A_{1g}$ mode due to out of plane vibrations of S atoms.[43] The Raman intensity map for the $A_{1g}$ mode for the sample T2 is shown in Figure 3a. The inset shows the optical micrograph of the Raman measurement area. Figures 3b and 3c show the peak position maps for $E^1_{2g}$ and $A_{1g}$ modes, respectively. It can be seen that the grain is quite uniform in Raman intensity and peak positions. Figure 3d shows Raman spectra acquired at four different spots on the grain (shown as dots in Figure 3a). The uniform intensity in the Raman map and similar Raman spectra indicate uniform crystallinity of the grain on sample T2. Figure 3e shows the $A_{1g}$ intensity map for the grain on sample T3, again with an optical micrograph in the inset. The grain morphology is similar to the one deposited on sample T2. The peak position distributions of $E^1_{2g}$ and $A_{1g}$ mode (Figure 3f and 3g, respectively) were found to be different at the center of the grain as compared to the edges. Figure 3h shows single point Raman spectra acquired at different locations on the grain (Figure 3e). A frequency difference of approximately 20 - 21 cm$^{-1}$ between the two prominent Raman modes indicates the formation of monolayer $MoS_2$ grains on samples T2 and T3. This demonstrates the uniformity of the $MoS_2$ phase synthesized at different growth temperatures. The map of frequency difference between the two modes further confirms the uniformity of the monolayer (Figure S3). However, in case of sample T3, the Raman modes exhibited a small yet observable blue-shift at the center as compared to the edges of the grain, with shift being more significant for the $E^1_{2g}$ mode than $A_{1g}$. Strain and doping have significant effects on the Raman modes of monolayer $MoS_2$. Chakraborty *et al*. have shown softening and broadening of the $A_{1g}$ mode with electron doping with no significant changes in the $E^1_{2g}$ mode.[44] This has been attributed to a stronger electron-phonon coupling of the $A_{1g}$ mode than of the $E^1_{2g}$ mode. On the other hand, strain is found to affect mostly the $E^1_{2g}$ mode with significant changes in its frequency and line-width or FWHM.[45,46] In our case, we observe a significant change in the $E^1_{2g}$ modes, which indicates that strain is responsible for the observed variations in Raman features across the grain.

Figure 4 shows representative PL measurements performed on the same grains shown in Figure 3a and 3e, where Raman measurements were carried out. Monolayer $MoS_2$ is a direct band gap material (1.8 eV – 1.9 eV) and it exhibits a strong PL near the direct gap at the K point of the Brillouin zone.[12] Due to the strong spin-orbit coupling in $MoS_2$, the valence band splitting leads to two excitations i.e. A and B excitons arising from the direct gap transitions between the maxima of the split valence bands and the minima of the conduction band. Therefore, a strong PL peak attests the formation of monolayer $MoS_2$ and corroborates the Raman results. Figure 4 shows the PL intensity maps and profiles of the A exciton peak for sample T2 and T3, respectively. The intensity is rather uniform for T2 (Figure 4a and 4b), while it is considerably higher at the center compared to the periphery of the grain for T3 (Figure 4d and 4e). The point PL spectra extracted from four different locations of the grains on sample T2 and T3 are shown in Figure 4c and 4f, respectively. They vary significantly at different locations of sample T3, confirming the Raman measurements. The A exciton peak exhibits observable changes of around 20 meV red-shift in peak position and a reduction in intensity towards edges of the grain. This behavior was consistently observed on other isolated grains on the same substrate, which appeared to have similar morphology (Figure S3). Moreover, the A exciton peak is blue-shifted and the peak intensity is higher, on an average, in case of sample T3 as compared to T2 (Figure

S5a). As a general observation, the much higher intensity of the A peaks than B peaks in the PL spectra of both the samples demonstrate the monolayer nature of the grains.[12] In the case of multilayer films, the intensity of the A and B peaks is nearly the same (Figure S5b). Kim *et al*. observed strong PL intensity in the center of $MoS_2$ grains grown in an excess S atmosphere.[35] However, those grains have the shape of truncated triangles in contrast to the full triangles observed in the present case. Moreover, we have synthesized $MoS_2$ under the same experimental conditions in the present work except a different deposition temperature. Such variations in PL features have been observed in CVD $MoS_2$ and have been attributed to local variations of strain in the isolated grains arising from the growth process, i.e. different thermal expansion coefficients of silica ($SiO_2$) and $MoS_2$.[47] However, the study was limited to a single growth temperature of 850 °C and strain variations on these samples were discussed. Nevertheless, given the range of temperatures for growing monolayer $MoS_2$ films, how the deposition temperature can affect the growth-induced strain is not very clear. In the present study, we are comparing the samples grown at different deposition temperatures and we find that strain variations become less prominent at lower deposition temperatures.

A detailed analysis of the acquired Raman and PL data was performed in order to understand the distinct optical properties of monolayer $MoS_2$ grains deposited at different temperatures. The schematic in Fig 5a depicts the locations from where single Raman and PL spectra were extracted. Locations 1 – 3 are near the corners of the triangular grain and location 4 is at the center of the grain. It is to be noted here that most of the reported single Raman and PL spectra in the literature are generally acquired at the center of the as-deposited grains. The peak positions of $E^1_{2g}$ and $A_{1g}$ modes acquired at different locations of the grains are plotted in Figure 5b. The frequency difference between the two modes is slightly higher at the corners (20.7 cm$^{-1}$) in the case of sample T3 and is nearly the same (20 cm$^{-1}$) at the center of the grains for both T2 and T3. This is also evident from the map of the frequency difference between the two modes (Figure S2). In the case of sample T2, the frequency difference is nearly the same at all the locations. Both the Raman modes are red-shifted at the corners of the grain in sample T3, with a maximum shift of around 1 cm$^{-1}$ for the $E^1_{2g}$ mode and 0.6 cm$^{-1}$ for the $A_{1g}$ mode. Such red-shifts of Raman modes of monolayer MoS$_2$ have been observed during strain tests, with the shift being more pronounced for the $E^1_{2g}$ mode.[45,48] However, no such shift is observed for sample T2. Figure 5c shows the variation of the full width at half maximum (FWHM) of $E^1_{2g}$ and $A_{1g}$ modes across the grains. Here, we did not observe a clear trend and FWHM values were more or less the same.

Above strain of about 1%, the $E^1_{2g}$ mode splits into two peaks due to the symmetry breaking of the crystal.[46] However, we did not observe a splitting of the $E^1_{2g}$ mode (Figure 3d and 3g), which suggests that strain levels are well below 1% in our samples. Rice *et al.* have reported a shift-rate of approximately -2.1 cm$^{-1}$ per % strain in the $E^1_{2g}$ mode of mechanically exfoliated monolayer MoS$_2$.[45] Taking this value as a first indication, we arrive at a strain of about 0.47% at the corners of the grain compared to the center of the grain on sample T3. Figure 5d shows the variation in the most intense PL peak corresponding to the A exciton across the grains on samples T2 and T3. The peaks at the edges appear at energies nearly 20 meV lower for both samples and approximately 30 meV lower in the center of the MoS$_2$ grains in sample T2 compared to T3. At the same time, there is a variation in the peak energy within the grains variations amounting to 20 meV in the case of T3 and 10 meV in the case of T2, with higher energy at the center of the grains than at the edges. Also, the FWHM of the PL peak is much higher in the case of T2. Such variations in band gap energy are attributed to non-uniform strain across the grains. To attest our present observations, we deposited another sample at a higher temperature of 750 °C and found significant variations in A exciton peak position across the grain (Figure S7). The peak is around 10 meV higher at the center of grain as compared to sample T3, indicating a higher residual strain in the sample grown at 750 °C.

We have performed *ab initio* density-functional theory (DFT) calculations of the electronic band structure of a MoS$_2$ monolayer undergoing different degrees of homogeneous strain in order to explain the experimental observations. Figure 6a shows the obtained band structures for the primitive cell depicted in Figure 6b, including spin-orbit coupling, and highlighting the $K_m$-$K_{v1}$ and $K_m$-$K_{v2}$ direct bandgaps, which correspond to A and B excitons, respectively. In Figure 6c, we show the variation of the direct band gaps with respect to the equilibrium case as a function of the Mo-Mo strain percentage (i.e., relative Mo-Mo distance with respect to the equilibrium value, while increasing/decreasing the lattice constant). Details on the calculations are included in the Methods section. The behavior for both the bandgaps is very similar and corresponds to a linear increase under the lattice compression, and a linear decrease in the case of tensile strain (roughly -120 meV/%). Considering this rate of change for the band energy, we estimate a strain variation of about 0.08% and 0.17% within the grains in the case of T2 and T3, respectively. This suggests that MoS$_2$ deposited at higher temperature has higher strain variation within the single grains. These values are well within the maximum strain value of 0.4% estimated from the variation in Raman modes across the grains.

Theoretical and experimental data from literature confirm our results: a decrease in band energy has been estimated to be -44 meV/% with an increase in uniaxial strain using density functional theory.[47] Conley *et al*. observed A peak redshifts at a rate of approximately 45 meV/%.[46] In addition, Michail *et al*. have proposed a methodology for optical detection of strain and doping inhomogeneities in exfoliated monolayer MoS$_2$.[49] Applying this method to our case, the maximum strain and carrier concentration is of the order of nearly 0.4 % and $1 \times 10^{13}$ cm$^{-2}$ for T2 and 0.5% and $0.9 \times 10^{13}$ cm$^{-2}$ for T3, respectively (Figure S5). Again, these values match those by by Michal *et al.*[49] for CVD MoS$_2$ on SiO$_2$ substrate, where strain varied from 0.2 to 0.3% and carrier concentration was around $1 \times 10^{13}$ cm$^{-2}$. Our own data in combination with these considerations strongly suggests that strain variation is the dominating factor rather than the carrier concentration in the CVD grown monolayer MoS$_2$.

It should be noted that uniaxial strain has been mostly proposed so far to explain the observed spectral changes in MoS$_2$. This may, however, not explain the growth-induced strain in monolayer MoS$_2$. Therefore, we have calculated the band energy shift of monolayer MoS$_2$ under homogeneous strain and provide the possible origin of strain by taking into account the adhesion of MoS$_2$ on silica (SiO$_2$) substrate.

Atomistic modeling can also provide information on the origin of the observed strain and its connection with the deposition temperature. The structure of native SiO$_2$/Si is that of an amorphous network, whose chemical termination depends on the experimental conditions. Spectroscopic data[50] and theoretical modeling[51] indicate that at 650-700 ºC the density of –OH is around 1.2-1.4 hydroxyls per 100 Å$^2$. At temperatures above 150 ºC some of these hydroxyls will turn into under-coordinated –O· (NBO or oxygen dangling bond) radical centers[51–54] and then act as anchoring centers for the MoS$_2$ monolayer. From atomistic DFT calculations (described in more detail in the supporting information), we can e.g. estimate the energetics of the process:

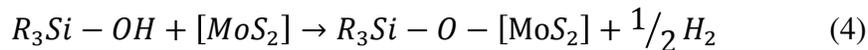

$$R_3Si - OH + [MoS_2] \rightarrow R_3Si - O - [MoS_2] + \tfrac{1}{2} H_2 \qquad (4)$$

The adhesion of a MoS$_2$ monolayer to an –O· center of SiO$_2$ is 1.17 eV based on our DFT calculations. Using the experimental values of free energy of H$_2$ at the given temperature and a reasonable H$_2$ pressure of 10$^{-4}$ atm,[55] the corresponding reaction free energies (ΔG) for Eq. (4) are predicted to 0.17 eV, 0.22 eV and 0.3 eV at 750 °C, 700 °C and 650 °C, respectively. The density of anchoring centers is thus predicted to be ≈14%, ≈6% and ≈2% of the hydroxyl density, and a corresponding adhesion energy per MoS$_2$ unit cell of 0.014 eV, 0.006 eV and 0.002 eV at 750 °C, 700 °C and 650 °C, respectively, is obtained. In the growth process of MoS$_2$ over SiO$_2$, formation of such anchoring centers will freeze the lattice parameter of the monolayer to the lattice parameter of the substrate at the growth temperature. When cooling down from the growth temperature to room temperature, a residual strain will then be present, determined by the difference in thermal expansion of MoS$_2$ monolayer and underlying substrate. We take the thermal expansion of MoS$_2$ monolayer from previous calculations to be around 7×10$^{-6}$ °C$^{-1}$ above room temperature,[56] while for the thermal expansion of the SiO$_2$/Si surface we take the average of the thermal expansion coefficient of silicon (≈3-4×10$^{-6}$ above room temperature)[57] and that of silica (≈10$^{-7}$ °C$^{-1}$, thus negligible). We thus obtain an estimate for residual strain in the MoS$_2$ monolayer of 0.23%, 0.21% and 0.18% that will produce – according to Figure 6c – a decrease in the band gap of 27.6 meV, 25 meV and 21 meV at 750 °C, 700 °C and 650 °C, respectively. We note that an estimated adhesion energy of 6 meV per MoS$_2$ unit cell at 700 °C or of 14 meV at 750 °C is sufficient to overcome the minor energy penalty associated with a uniform strain of 0.2%, i.e., 2 meV per unit cell as predicted by our DFT approach (such small value can be explained on the basis of the 2D character of MoS$_2$[56,58]). At variance, our estimated adhesion energy of 2 meV per MoS$_2$ unit cell at 650 °C is at the boundary between anchoring and release at room temperature, in agreement with the experimental observation that strain at 650 °C is smaller than that predicted in the presence of resilient anchoring. In these considerations, we have ignored the increase in the kinetic rate of dehydrogenation and anchoring upon temperature that can also contribute to disfavor anchoring at 650 ºC.

In summary, we have studied the effect of deposition temperature on surface morphology and optical properties of CVD grown monolayer $MoS_2$. It is found that surface morphology, deposition behavior and optical properties of monolayer $MoS_2$ are significantly different for different growth temperatures. The strongest PL peak exhibited variations in intensity and band energy positions shifts of 10 to 20 meV with increasing deposition temperatures from 650 °C to 700 °C. Isolated $MoS_2$ grains showed variations in the band energies across the films, with more striking variations observed for $MoS_2$ deposited at 700 °C. Further verification of these observations was done by depositing monolayer MoS2 at 750 °C. DFT calculations confirm that such variations can result from different levels of residual strain in as-deposited $MoS_2$, with smaller strain levels observed at lower deposition temperatures. An atomistic model is introduced to explain the growth temperature dependent strain within individual grains. The present study demonstrates the important role of substrate temperature on the properties of CVD monolayer $MoS_2$ and, at the same time, opens opportunities to tune $MoS_2$ properties during growth through strain engineering. These findings and the theoretical models can also be applied to other 2D TMD materials and growth substrates.

**Experimental Section**

*MoS$_2$ Synthesis.* Monolayer MoS$_2$ films were deposited using MoO$_3$ and S precursors in a CVD reactor comprising of a two inch quartz tube. We used 325 nm SiO$_2$/Si substrates, which were cleaned using acetone and isopropyl alcohol before loading in the reactor. The substrates were kept upside down on an alumina boat containing MoO$_3$ precursor, located at the center of the furnace. The furnace was then heated to the desired deposition temperatures. The S powder was kept in another alumina boat at the upstream end of the tube. During all the experiments, S was kept at a temperature around 150 °C using an independent heating apparatus. This was done in order to have the similar S content during all the experiments. It is to be noted here that the substrates and MoO$_3$ precursor were at the same temperature due to the deposition configuration used in the present study. The depositions were carried out for 10 mins. at atmospheric pressure under a constant argon (Ar) gas flow of 20 sccm. After the deposition, the furnace was cooled down naturally to room temperature under a constant flow of Ar gas.

*Morphological and Optical Characterization.* As deposited samples were characterized using optical microscope and scanning electron microscope (Helios Nanolab 600) for morphological studies. Detailed Raman and PL studies were carried out using a confocal Raman microscope (WITec Alpha 300R) equipped with a 532 nm wavelength laser. The signal was collected through a 100× objective, dispersed with an 1800 g/mm grating for Raman and 300 g/mm grating for PL measurements. The signal was detected using a Peltier cooled charge-coupled device. During Raman and PL measurements, the laser power was kept at 0.5 mW and was measured using a power meter (ThorLabs).

*DFT calculations*. We have performed DFT calculations using Quantum Espresso package,[59] including plane wave basis set, a gradient-corrected exchange correlation functional (Perdew-Burke-Ernzerhof (PBE))[60] and ultrasoft pseudopotentials (US-PPs)[61].

In order to compute the bandgap variations as a function of the strain, we have carried out relaxation calculations of $MoS_2$ monolayer, considering different systems, while varying the lattice parameter '$a$' (Figure 6b). Each relaxation has been followed by single point electronic calculations (i.e., K point) performed at each optimized geometry. A hexagonal cell has been considered, with an initial lattice parameter of $a$ = 3.185 Å, and considering periodic conditions in the x and y directions. A vacuum region of about 25 Å in the z direction has been taken into account to minimize the interaction between adjacent image cells. An energy cut-off of 60 Ry has been considered for the selection of the plane-wave basis set for describing the wave function and 600 Ry as the cutoff for describing the electron density. The Brillouin zone (BZ) is sampled at $k_x$=$k_y$=12 and $k_z$=1. Spin-orbit coupling (SOC) has been included, considering fully relativistic approximation.

**Supporting Information**

Supporting Information is available from the Wiley Online Library or from the author.

* **Corresponding author:** max.lemme@uni-siegen.de , skataria2k2@gmail.com

ACKNOWLEDGEMENTS

Support from the European Research Council (ERC, InteGrade, 307311), the German Ministry of Education and Research (BMBF, NanoGraM, 03XP0006C) and the German Research Foundation (DFG LE 2440/1-2 and LE 2440/2-1) is gratefully acknowledged. G.F. and G.I. gratefully acknowledge the Graphene Flagship (696656).

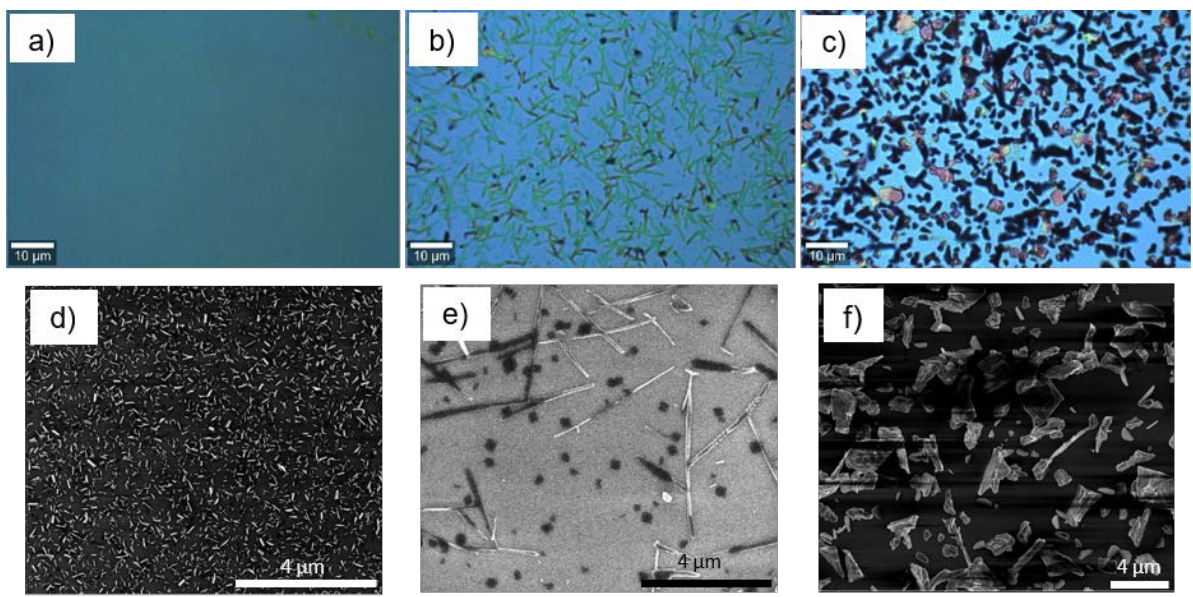

**Figure 1.** Surface morphology of the samples deposited at positions directly above the precursor. Optical micrograph of sample T1 (a), T2 (b) and T3 (c). The micrographs are enhanced for better clarity. A difference in morphology can be observed clearly. Corresponding scanning electron micrographs of samples T1 (d), T2 (e) and T3 (f), deposited at 600 °C, 650 °C and 700 °C, respectively.

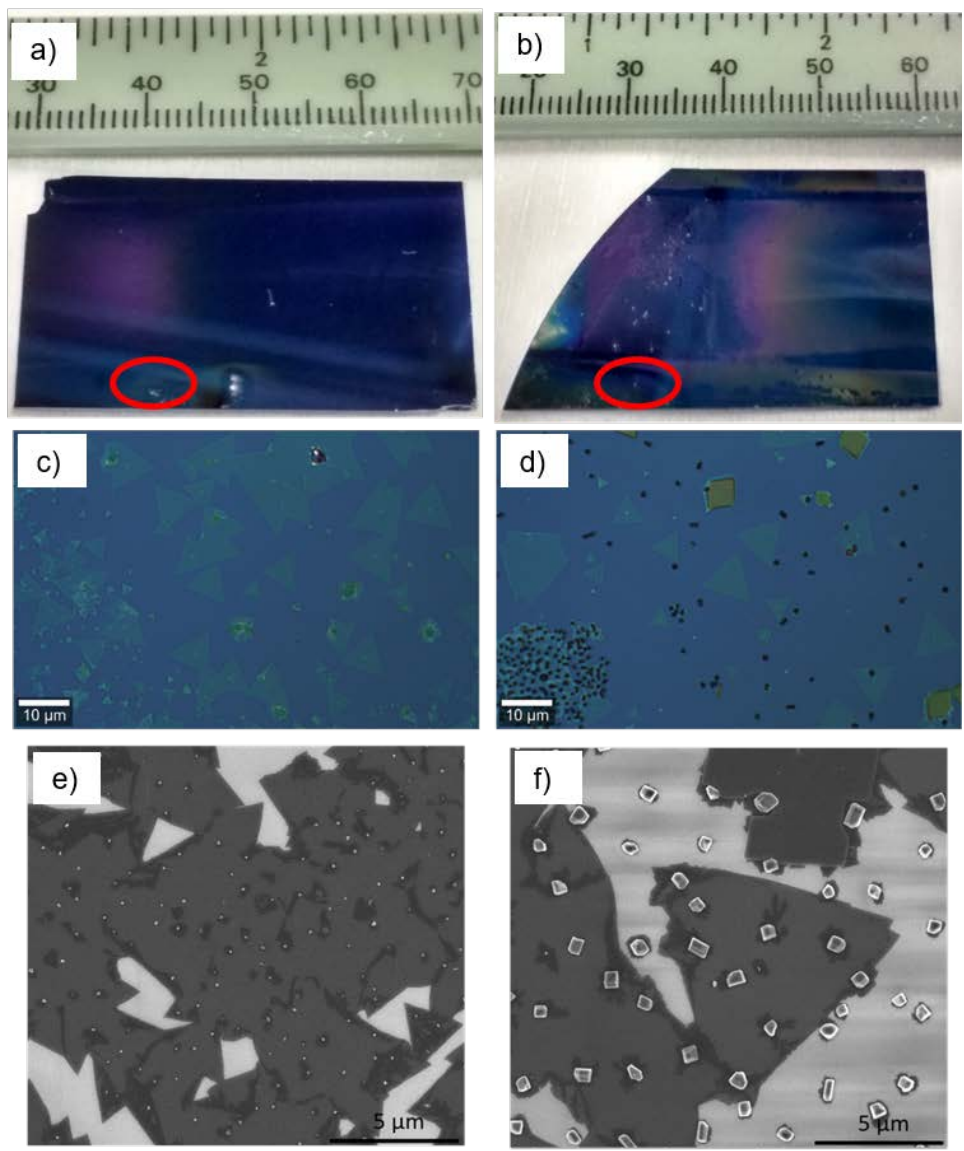

**Figure 2.** Morphology of isolated monolayer MoS$_2$ deposited near the substrate edges. Photographs of sample T2 (a) and T3 (b) depicting the deposition contour along the substrate length. The direction of gas flow during deposition was from left to right. The red oval marks the position of further optical and SEM studies on the respective samples. Optical micrographs of sample T2 (c) and T3 (d). SEM micrographs of sample T2 (e) and T3 (f). Different shapes of nanoparticles are observed in addition to grains of MoS$_2$ with increased density and size for sample T3.

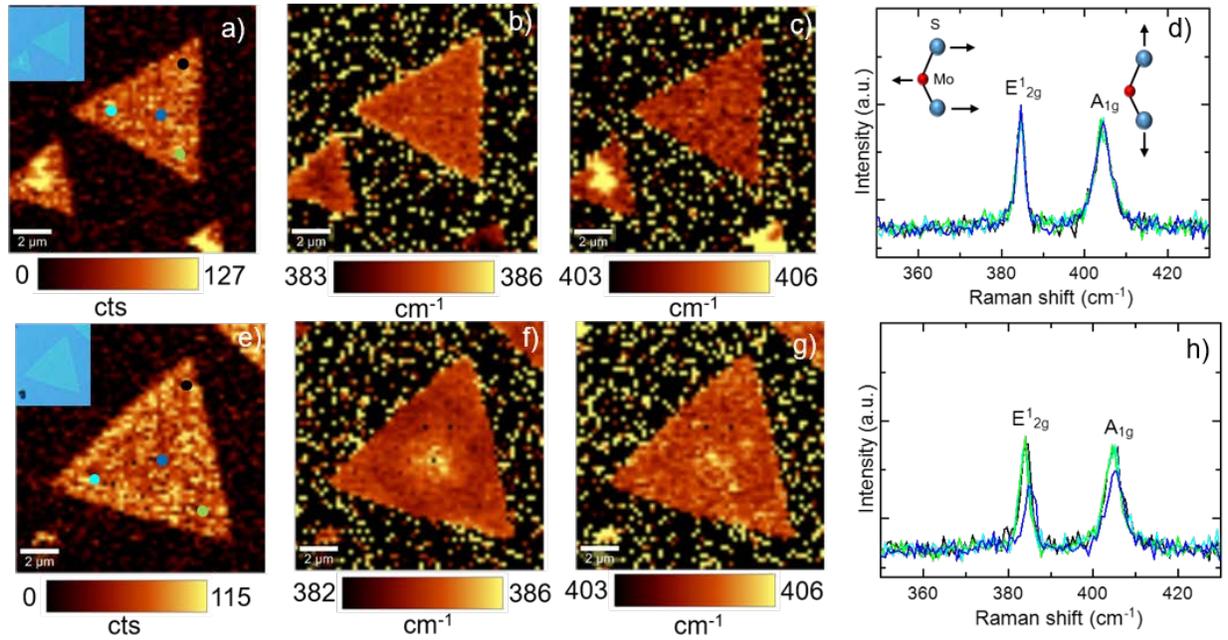

**Figure 3.** Raman spectroscopy of as-deposited $MoS_2$ samples. Two-dimensional Raman intensity maps of the $A_{1g}$ peak (a), Raman peak position maps for the $E^1_{2g}$ and the $A_{1g}$ mode (b, c), and point spectra extracted at four different locations on the $MoS_2$ grain for sample T2. The same data are shown for sample T3 in (e)-(h). The insets in (a) and (e) show the optical micrograph of the respective grains where Raman measurements were performed. The dots in (a) and (e) show the locations from where point spectra are extracted (shown in (d) and (h), respectively).

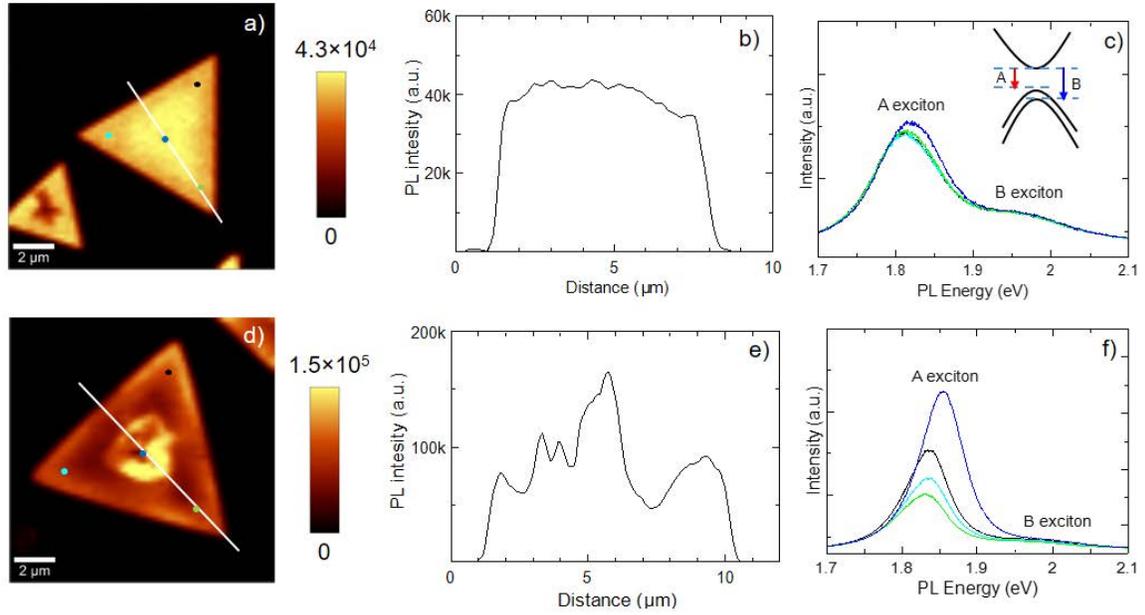

**Figure 4.** Photoluminescence (PL) study on isolated $MoS_2$ grains. PL intensity map of the A exciton peak (a), the intensity profile (b) along the white line in (a), and PL spectra (c) extracted from locations represented by dots of same color in (a) for sample T2. The schematic in (c) represents the A and B excitons from conduction band to splitted valence bands. PL intensity map (d), intensity profile of the A exciton peak (e) along the white line in (d), and point spectra (f) acquired at positions shown by dots of the same color in (d) for sample T3.

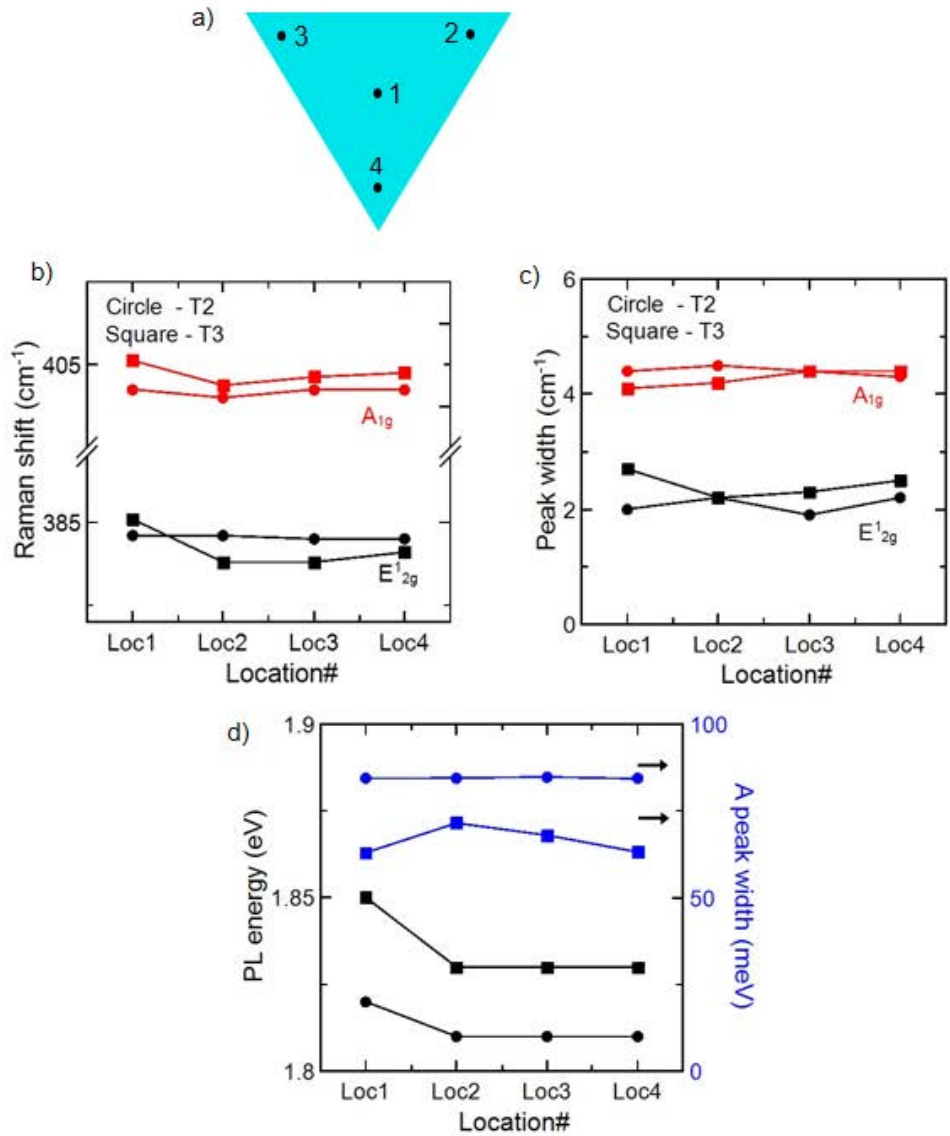

**Figure 5.** Raman and PL data analysis. (a) Schematic of triangular grain showing the locations from where Raman and PL point spectra were obtained. The individual spectra used for analysis are shown in Figs. 3 and 4. (b) Raman peak positions for $E^1_{2g}$ and $A_{1g}$ modes at different locations. (c) FWHM of $E^1_{2g}$ and $A_{1g}$ peaks and, (d) PL energy and peak width of the A exciton peaks.

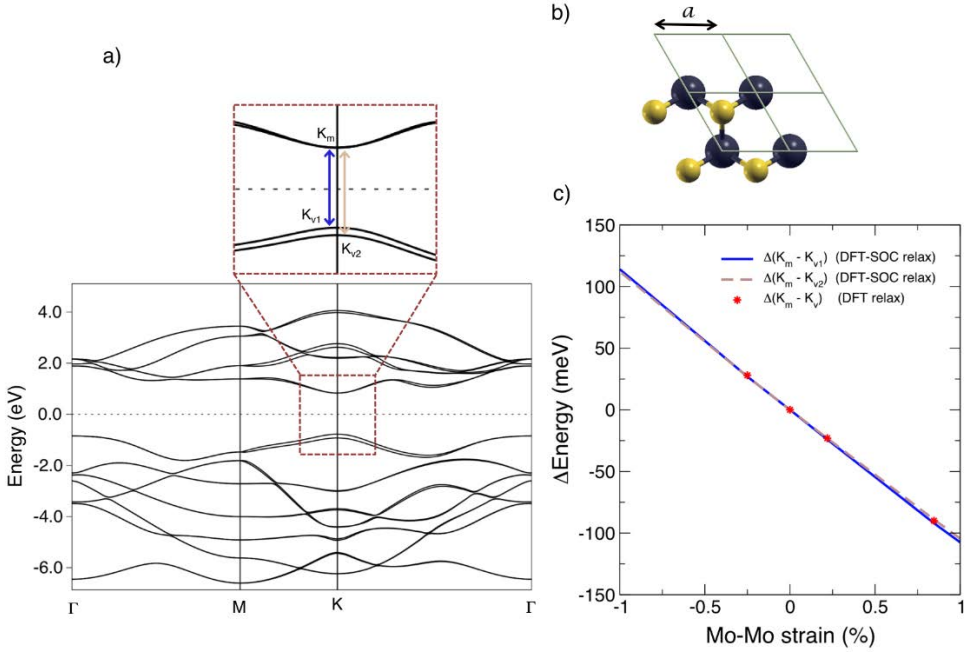

**Figure 6.** Calculated band structure of monolayer MoS$_2$: a zoom in correspondence of the K-point is shown in the inset (a). Schematic of the primitive cell considered for the calculations (b). Variations of the band-gaps highlighted in (a), for different strains, including and excluding spin-orbit coupling (SOC) (c).

# Supporting information

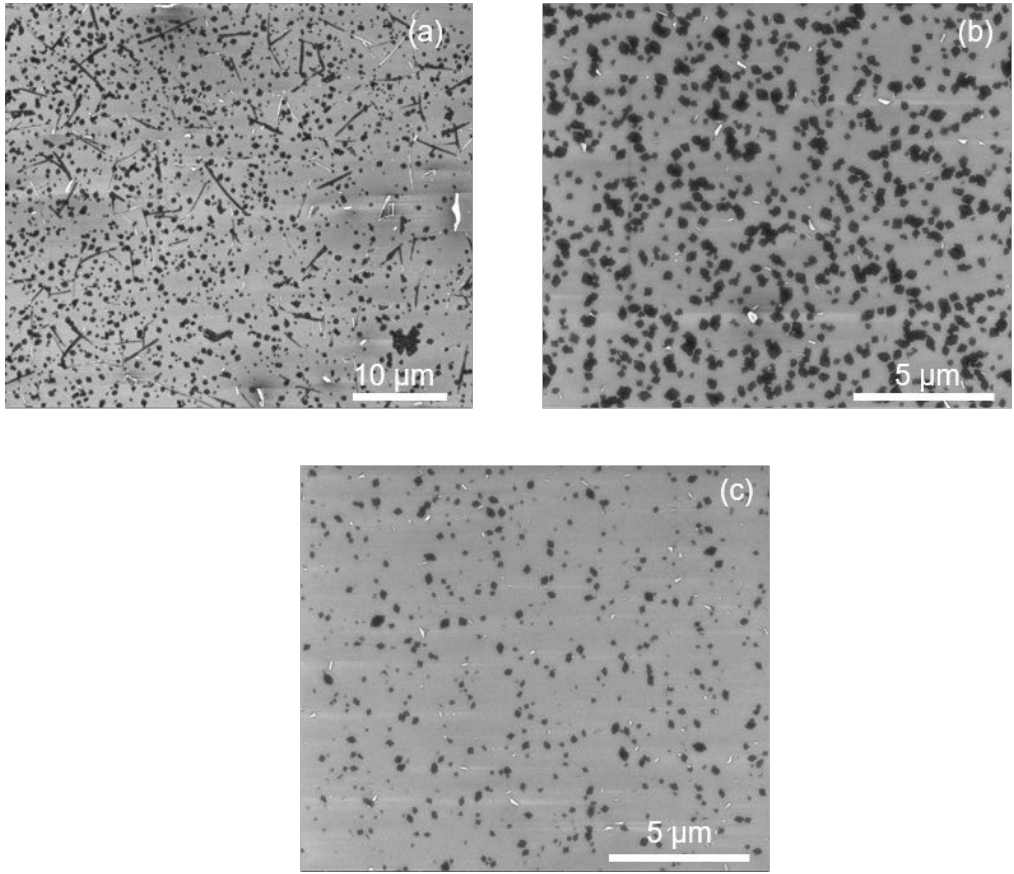

**Figure S1.** SEM micrographs showing the deposition along the substrate length in downstream direction from (a) to (c).

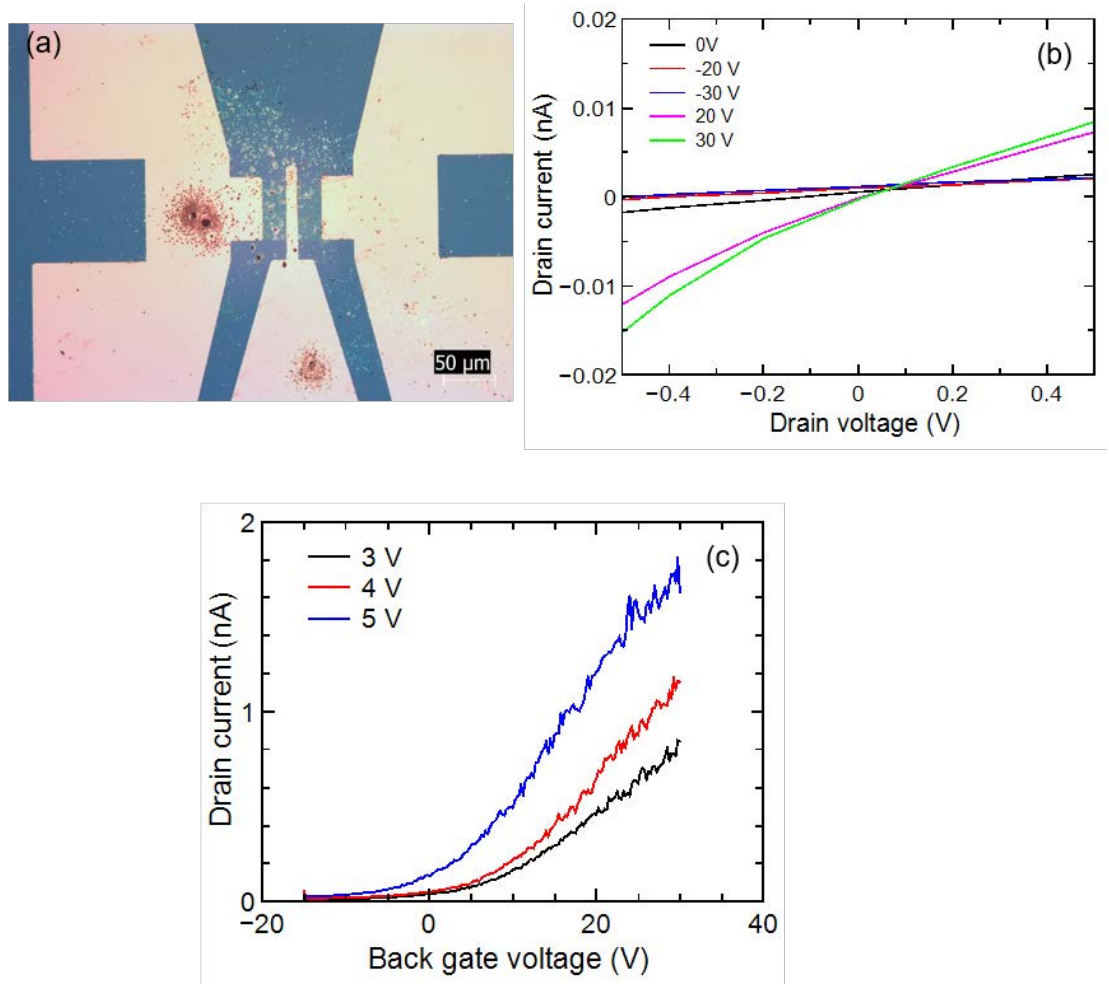

**Figure S2.** (a) Optical micrograph of a back-gated single layer $MoS_2$ device. The continuous single layer $MoS_2$ grown at a temperature of 700 °C was transferred to a 325 nm $SiO_2$/Si substrate. Thermally evaporated Cr (5nm)/Au (100 nm) stack was used as source and drain contacts. (b) Output and (c) transfer characteristics of $MoS_2$ transistor, at various source-drain voltages, depicting the n-type behavior of the channel.

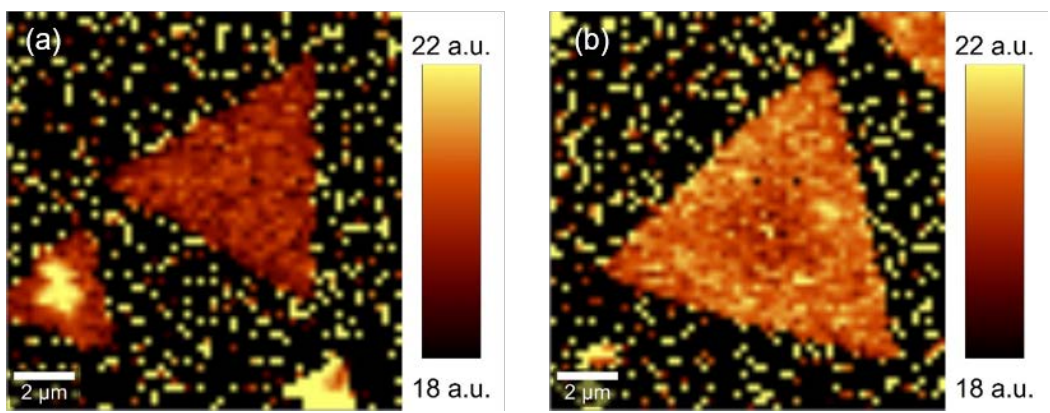

**Figure S3.** Map of frequency difference between $E^1_{2g}$ and $A_{1g}$ mode for sample (a) T2 (650 °C) and (b) T3 (700 °C).

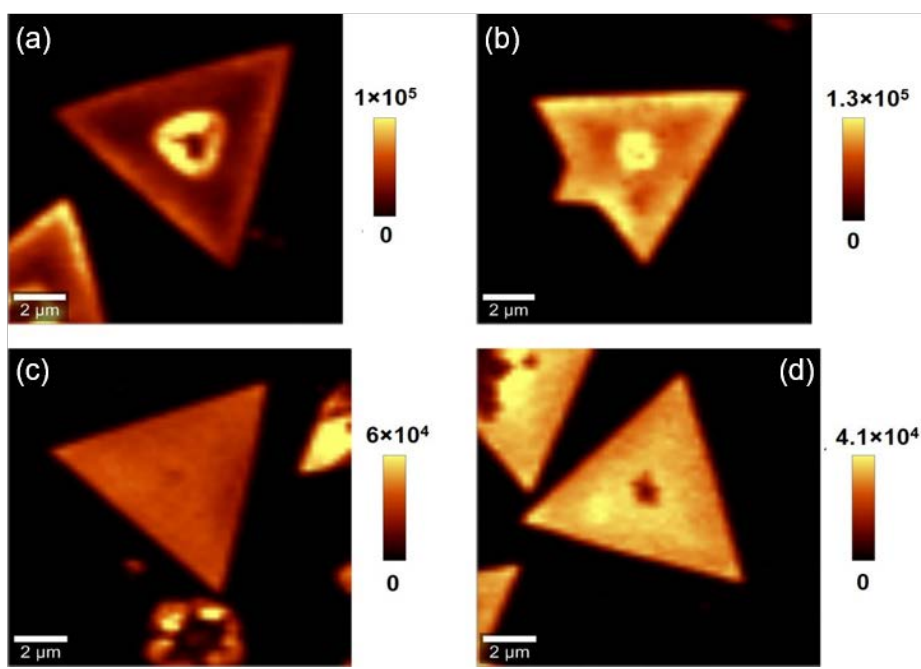

**Figure S4.** PL intensity map of A exciton peak for different MoS$_2$ grains on sample T3 (700 °C) ((a) and (b)), and sample T2 (650 °C) ((c) and (d)). It can be seen that PL intensity is much higher at the center of flakes in case of sample T3.

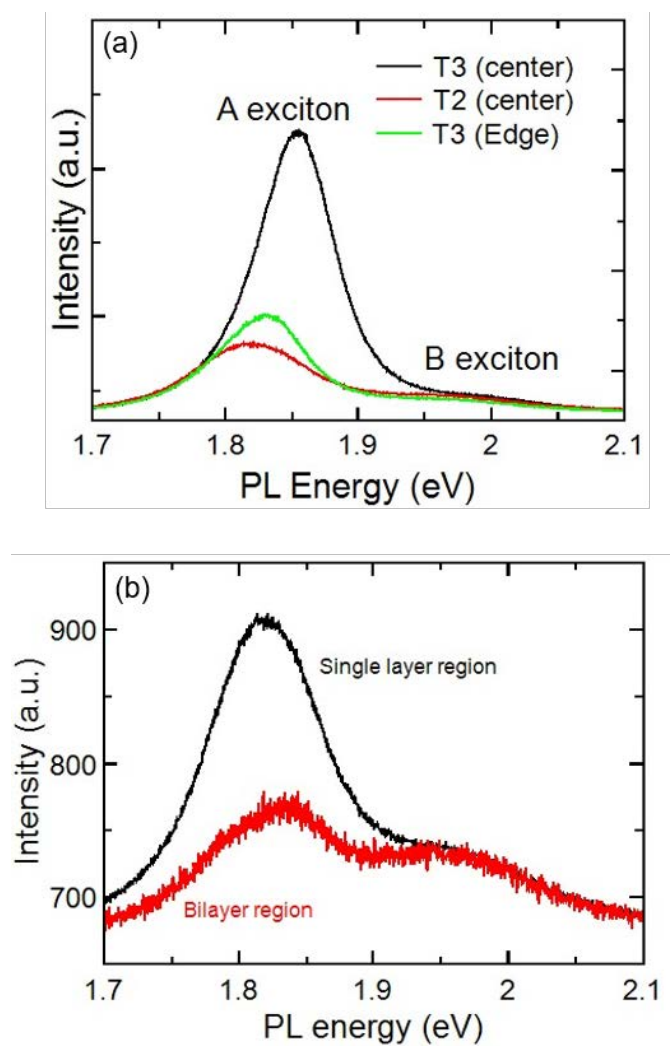

**Figure S5.** (a) Comparison of PL spectra of single layer $MoS_2$ grown at samples T2 (650 °C) and T3 (700 °C). (b) PL spectra of single and bilayer region on sample T2.

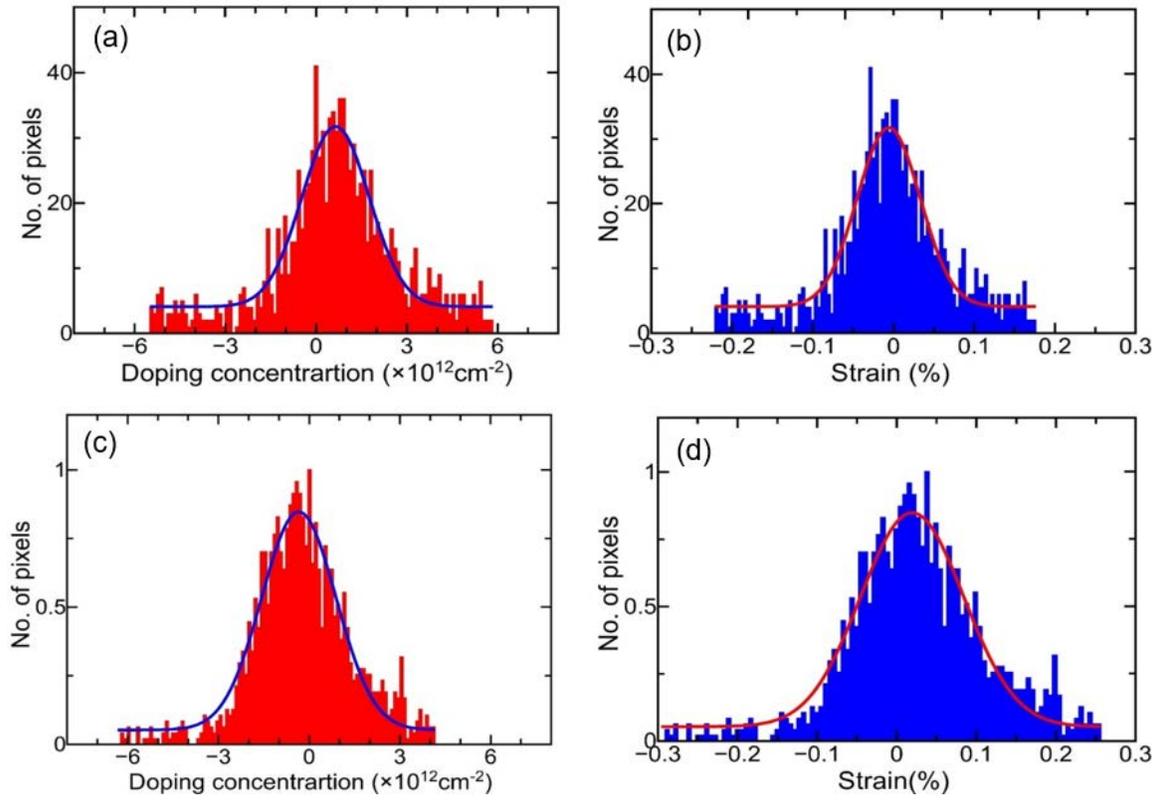

**Figure S6.** Doping concentration and strain distribution for sample T2 (650 °C) ((a), (b)) and T3 (700 °C) ((c), (d)). The analysis was performed using the methodology presented by Michail *et al*.[1] We have used Raman mapping data shown in Fig. 3(a) and 3(e) with mean position values of $E^1_{2g}$ and $A_{1g}$ as 383.15 cm$^{-1}$ and 404.75 cm$^{-1}$, respectively.

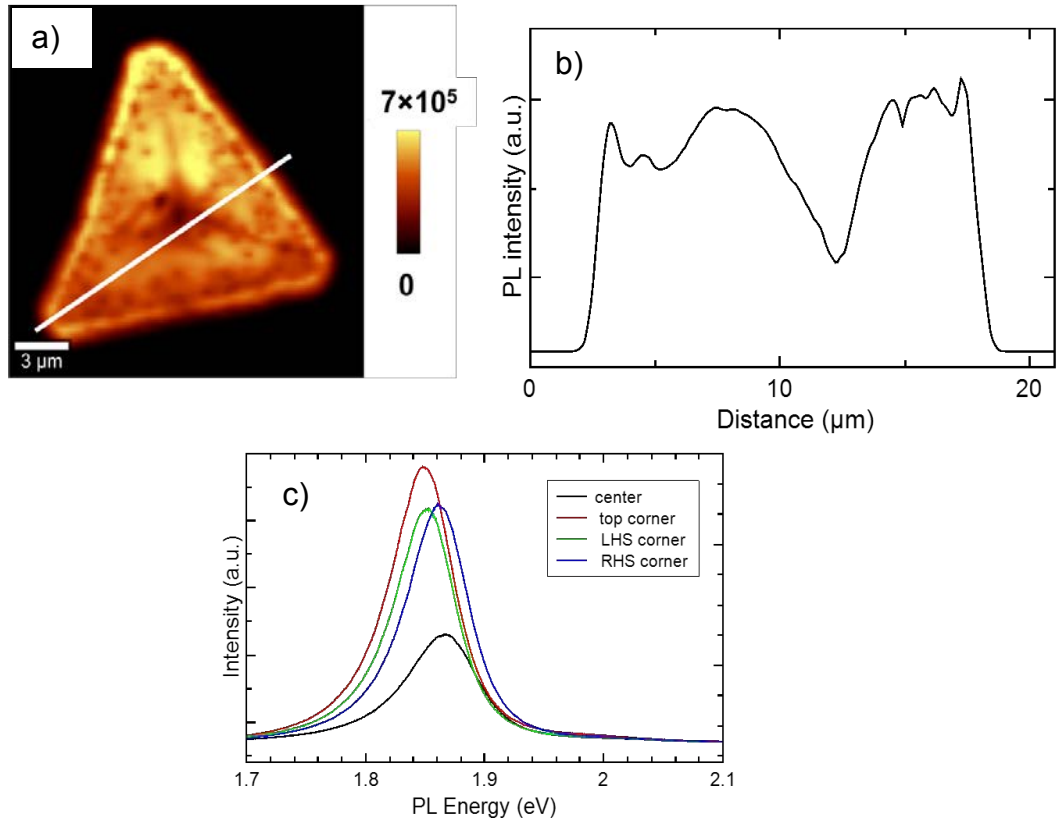

**Figure S7**.: Photoluminescence (PL) studies on a sample grown at 750 °C. PL map of A exciton (a). PL intensity profile along the white line in (a) across the isolated grain (b). Single PL spectra extracted from the described locations (c). We observed a red-shift of around 10 meV in A exciton peak position in the center of grain as compared to the edges. Also, the A exciton peak in the center is at 1.86 eV which is around 10 meV higher than that observed in sample T3.

**Geometry optimization of suspended flake**

Geometry optimization of suspended flake has been performed, in order to investigate tensile/compressive strain in $MoS_2$. A suspended triangular structure of about 1.9 nm per side has been considered, taking into account two configurations, depending on the $MoS_2$ edge termination i.e., with zig-zag geometry and S atoms at edges, and S-edge termination as proposed by Huang *et al.*[2] (Figure S7a and S8a, respectively). As can be seen, in both structures a tensile strain is observed in close proximity of the borders (0.1% and 5.2%, in Figure S7b and Figure S8b, respectively), while the same equilibrium distance in bulk monolayer $MoS_2$ is observed in atoms 1 nm far from the borders. This may suggest that the strain experimentally observed in micro-meter scale flakes, is probably due to interactions with $SiO_2$ substrate and/or defects, which may increase with temperature, inducing changes in the lattice parameters.[3,4]

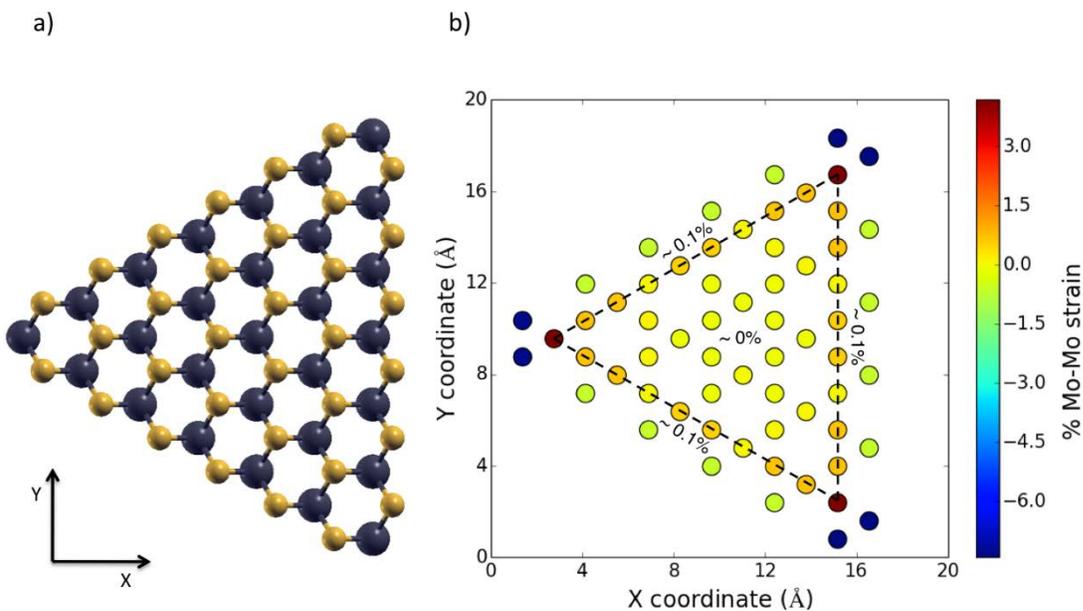

**Figure S8:** (a) Simulated structures with zig-zag geometry at the edges; (b) colormap of the Mo-Mo distance computed by means of DFT calculations.

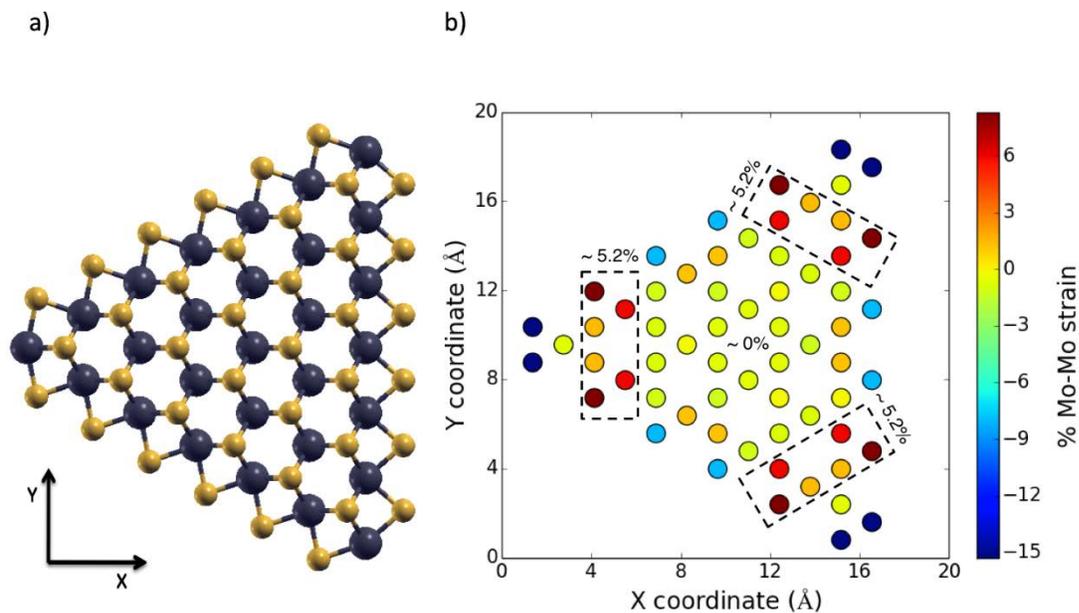

**Figure S9:** (a) Simulated structures with edge geometry as in;[5] (b) colormap of the Mo-Mo distance computed by means of DFT calculations.

**Simulation of the interaction of Si(OH)$_3$O- with a MoS$_2$ monolayer**

We have performed DFT simulations in order to obtain the adhesion energy of a MoS$_2$ monolayer and an NBO center on silica: -Si-O$^-$. Adhesion energy was calculated as the difference between the total energy of the system and the energy of each single component in the interaction configurations. In order to obtain the total energies, relaxation and single point electronic calculations were performed, considering one molecule of Si(OH)$_3$O$^-$ placed on an MoS$_2$ monolayer (as shown in Figure S9). In this case, the cell parameter is a = 12.74 Å and a vacuum region of about 25 Å in the z direction has been taken into account. An energy cut-off of 40 R$_y$ is used for the selection of the plane-wave basis set for describing the wave function and 400 R$_y$ as the cutoff for describing the electron density. The Brillouin zone (BZ) is sampled at k$_x$=k$_y$=3 and k$_z$=1. Dispersion effects (van der Waals corrections[6]) were included.

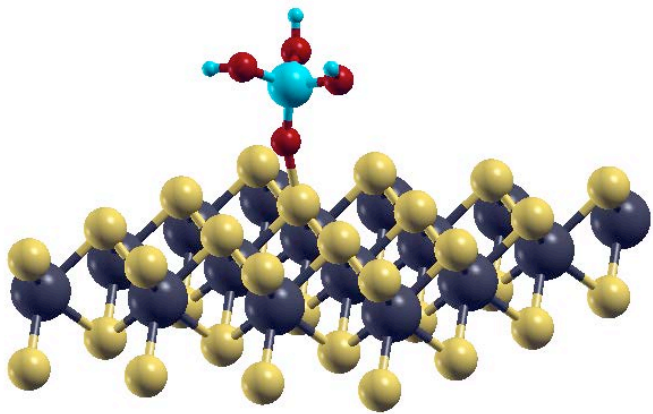

**Figure S10:** Simulated structure for the investigation of the interaction between MoS$_2$ monolayer and Si(OH)$_3$O$^-$.